\documentclass{amsart}
\usepackage{amssymb}
\usepackage{amsmath}
\usepackage{amsfonts}
\usepackage{graphicx}
\usepackage{epstopdf}
\usepackage[vcentermath]{youngtab}
\usepackage{float}
\usepackage{placeins}

\newcommand{\beq}{\begin{equation}}
\newcommand{\eeq}{\end{equation}}

\begin{document}

\begin{center}
{\Large\bf The Casimir eigenvalues on $ad^{\otimes k}$ of SU(N) are linear on N
}\\
\vspace*{1 cm}

{\large  R.L.Mkrtchyan 
}

\vspace*{0.5 cm}

{\small\it Yerevan Physics Institute, 2 Alikhanian Br. Str., 0036 Yerevan, Armenia}

{\small\it E-mail: mrl55@list.ru}

\end{center}

 {\bf Abstract.} 
	
We consider eigenvalues of the Casimir operator on the naturally defined \textit{stable sequences} of representations  of $su(N)$ algebra and prove that eigenvalues are linear over $N$ iff  $\lambda_1+2\lambda_2+...+k\lambda_k=\lambda_{N-1}+2\lambda_{N-2}+...+k\lambda_{N-k}$, where $\lambda_i$ are Dynkin labels, and $\lambda_i=0$ for  $k<i<N-k$, with fixed $k$. These representations are exactly those which appear in the decomposition of  $ad(su(N))^{\otimes k}$, therefore this linearity admits the presentation of eigenvalues in the universal, in Vogel's sense, form, and supports the hypothesis of universal decomposition of $ad^{\otimes k}$  into Casimir eigenspaces. 
	

\section{Eigenvalues of Casimir operator}

 Let $X^a$ be generators of any simple Lie algebra in some irreducible representation. We can define corresponding Casimir operator with the help of an invariant metric $g_{ab}$:

\begin{eqnarray}
	C=g_{ab}X^aX^b
\end{eqnarray}

Invariant metrics $g_{ab}$ in simple Lie algebras are unique up to normalization, which we choose to be the minimal one, where the squares of the long roots of the algebra are equal to 2. 

There is a general formula for the eigenvalue of the Casimir operator on highest weight representations of simple Lie algebras. Denoting by $\lambda$ the highest weight vector, the Casimir eigenvalue on the corresponding representation is (see e.g. \cite{BR,Difr,GG}):

\begin{eqnarray}\label{Ceig}
	C=(\lambda,\lambda)+2(\lambda,\rho)
\end{eqnarray}
where $\rho$ is the Weyl vector in the root space, the half-sum of all positive roots of a given algebra, and the scalar product is with abovementioned metric.

One can calculate the  eigenvalue of Casimir operator on, say, the adjoint representation. It is given in our normalization by $2h^\vee$, where $h^\vee$ is the dual Coxeter number of the given algebra. Below we consider the algebra $su(N)$, for which $h^\vee=N$. 

For the $su(N)$ algebra we introduce Dynkin labels which are non-negative integers, coefficients in the decomposition of $\lambda$ in the basis of fundamental weights:  $\lambda=\lambda_1 \omega_1+ ...+ \lambda_{N-1} \omega_{N-1}$, with standard enumeration of simple roots of $su(N)$.

For the fundamental representation (i.e. Dynkin labels (1,0,0,0...)), the eigenvalue of Casimir is equal to  $N-\frac{1}{N}$; for the symmetric tensors representation (2,0,0,0...) it is $2N-\frac{4}{N}+2$; for the antisymmetric tensors (0,1,0,0...) it is $2N-\frac{4}{N}-2$; for the adjoint representation (1,0...0,1), as mentioned, $2N$, etc. 

The question which we would like to answer is: for which representations is the dependence of Casimir's eigenvalue on $N$ linear, i.e. there are no terms like $1/N$ in the above examples. Actually, as we shall see below, the general form of eigenvalues is as above - there are terms of order $N$, $1/N$, and constant. 

\section{Stable sequences of representations of $su(N)$}

An important subtlety is the very definition of "$N$-dependence" of eigenvalues, which remains disguised in the previous examples. Indeed, as we vary $N$, we have to define what sequence of representation we are considering. E.g. if we start from the fundamental representation of $su(3)$, with Dynkin labels (1,0), we can extend it to higher $N$ by adding zeros at the right end of the row of labels, keeping the representation the fundamental one for all $N$. But when we start with the adjoint representation (1,1), we add zeros between the two labels 1, to keep the representation the adjoint one for $su(N)$ with any $N$. Evidently, second example allows more general representations. 

So, we call the sequence of representations of $su(N)$, $N=N_0, N_0+1,...$ the \textit{stable sequence} if there exist some $N_1$ and $k$, with $2k<N_1$, such that  for all $N>N_1$: 

1) the first $k$  Dynkin labels of representations don't depend  $N$ 

2) the last $k$ Dynkin labels don't depend on $N$, and 

3) all other Dynkin labels are zero. 

In other words, we extend any representation of given $su(N_1)$ for larger $N$ by simply adding zeros "in the middle" of the row of Dynkin labels, between possibly non-zero first $k$ and last $k$ labels. 

With this definition, one can ask on dependence  of $N$  of different quantities, say dimensions, Casimir eigenvalues, etc., implying its value on the $N$-th ($N > N_1$) element of the stable sequence of representations. 

\section{Casimir eigenvalues for stable sequences}
The values of Casimir for a given representation can be calculated directly, via the matrix $F$ of scalar products of the fundamental weights:

\begin{eqnarray}
	F_{ij}=(\omega_i,\omega_j)
\end{eqnarray}
It is clearly sufficient for the calculation of the first term in (\ref{Ceig}), and also is sufficient for the second term, if we take  into account the other representation of Weyl vector:

\begin{eqnarray}
	\rho=\omega_1+...+\omega_{N-1}
\end{eqnarray}
as the sum of all fundamental weights.

One can give an explicit form of the matrix $F$ for $su(N)$  (\cite{Difr} p. 541, \cite{GG} (5.5.4)):

\begin{eqnarray}
F_{ij}=\frac{min(i,j)(N-max(i,j))}{N}=min(i,j)-\frac{ij}{N}
\end{eqnarray}

In a visualized form (appropriate for $k=4$) $F$ is:

{\scriptsize
	
\begin{eqnarray*}
\frac{1}{N}\left(
\begin{array}{ccccccccc}
	N-1 & N-2 & N-3 & N-4 & . & 4 & 3 & 2 & 1 \\
	N-2 & 2 (N-2) & 2 (N-3) & 2 (N-4) & . & 8 & 6 & 4 & 2 \\
	N-3 & 2 (N-3) & 3 (N-3) & 3 (N-4) & . & 12 & 9 & 6 & 3 \\
	N-4 & 2 (N-4) & 3 (N-4) & 4 (N-4) & . & 16 & 12 & 8 & 4 \\
	. & . & . & . & . & . & . & . & . \\
	4 & 8 & 12 & 16 & . & 4 (N-4) & 3 (N-4) & 2 (N-4) & N-4 \\
	3 & 6 & 9 & 12 & . & 3 (N-4) & 3 (N-3) & 2 (N-3) & N-3 \\
	2 & 4 & 6 & 8 & . & 2 (N-4) & 2 (N-3) & 2 (N-2) & N-2 \\
	1 & 2 & 3 & 4 & . & N-4 & N-3 & N-2 & N-1 
\end{array}
\right)
\end{eqnarray*}

}

One can calculate the sum of entries of $F$ in each row, necessary for the second term in (\ref{Ceig}). The result is

\begin{eqnarray}
R_i=	(\omega_i,\rho) = \sum_{j=1}^{N-1}F_{ij}=\frac{i(N-i)}{2}
\end{eqnarray}

So, according to the above, we have the Dynkin labels $\lambda_i$, the non-negative integers, equal to zero for $k <i< N-k$ for some fixed $k$. The labels 
$\lambda_i, i\leq k$ are independent of $N$, so are the last labels $\lambda_{N-i}, i=1,...,k$, despite the notation. To stress this last fact we denote 
$\lambda_{N-i}=\tau_i, i=1,...k$. 

The eigenvalue of the Casimir on such representations is
\begin{eqnarray}
	C=(\lambda,\lambda)+2(\lambda,\rho) = \\
	\sum_{i,j=1}^{k}\left( \lambda_i \lambda_j F_{ij}+ \tau_i \tau_j F_{N-i,N-j}+ \lambda_i \tau_j F_{i,N-j}+ \tau_i \lambda_j F_{N-i,j}\right) + \\
	2 \sum_{i=1}^{k} (R_i \lambda_i +R_{N-i}\tau_i)
	\end{eqnarray}

The sums are:

\begin{eqnarray}
	\sum_{i,j=1}^{k}\lambda_i \lambda_j F_{ij}=\sum_{i,j=1}^{k} min(i,j) \lambda_i \lambda_j - \frac{1}{N} \left( \sum_{i=1}^{k} i\lambda_i \right)^2 \\
	\sum_{i,j=1}^{k}\tau_i \tau_j F_{N-i,N-j}=\sum_{i,j=1}^{k} min(i,j) \tau_i \tau_j - \frac{1}{N} \left( \sum_{i=1}^{k} i\tau_i \right)^2 \\
	\sum_{i,j=1}^{k}\left( \lambda_i \tau_j F_{i,N-j}+ \tau_i \lambda_j F_{N-i,j}\right) =\frac{2}{N} \left( \sum_{i=1}^{k}i\lambda_i \right) \left( \sum_{j=1}^{k} j\tau_j \right)  \\
	2 \sum_{i=1}^{k} (R_i \lambda_i +R_{N-i}\tau_i)=N \sum_{i=1}^{k}  \left(  i\lambda_i+i\tau_i   \right) - \sum_{i=1}^{k} \left(  i^2\lambda_i+i^2\tau_i   \right)
\end{eqnarray}

Altogether we have for the eigenvalues of Casimir in terms of the Dynkin labels: 

\begin{eqnarray}
C=	N \sum_{i=1}^{k}  \left(  i\lambda_i+  i\tau_i   \right) +    \\
\sum_{i,j=1}^{k} min(i,j) \lambda_i \lambda_j+\sum_{i,j=1}^{k} min(i,j) \tau_i \tau_j  - \sum_{i=1}^{k} \left(  i^2\lambda_i+i^2\tau_i   \right) - \\
\frac{1}{N} \left( \sum_{i=1}^{k} ( i\lambda_i -i\tau_i ) \right)^2
\end{eqnarray}

We see that the eigenvalues are linear over $N$, i.e. $1/N$ term is absent, if and only if 

\begin{eqnarray} \label{Main1}
	\sum_{i=1}^{k} ( i\lambda_i -i\tau_i )=0
\end{eqnarray}

or, in the standard notations,

\begin{eqnarray} \label{Main2}
\lambda_1+2\lambda_2+...+k\lambda_k=\lambda_{N-1}+2\lambda_{N-2}+...+k\lambda_{N-k}
\end{eqnarray}

Note that despite being quadratic in $\lambda, \tau$, the coefficient of $1/N$ appears to be a perfect square, which is why the condition for it to vanish becomes linear. 

This condition has a simple interpretation. The Young diagrams in stable sequences have two parts: the first  is described by the labels $\tau$ and consists of columns of height "of order $N$", more precisely their complement to length $N$ columns is the Young diagram with $\tau_i$ columns of heights $i\leq k$; and the second part is another diagram, described by labels $\lambda_i$, also with heights $i\leq k$. Then the condition (\ref{Main2}) means that areas of these two Young diagrams are equal.

\section{Conclusion. Universal decompositions of powers of adjoint}

Consider the decomposition of the $k$-th power of the adjoint representation of a simple Lie algebra into irreducible representations. Such decompositions are considered in a number of papers \cite{Del,V0,Cohen,LM}, also recently \cite{AIKM}, particularly  in connection with Vogel's universality. The adjoint representation of $su(N)$ can be realized in the space of second rank tensors with one "upper" and one "lower" indices (i.e. with one index from the fundamental representation and the other one from the conjugated one), with the zero trace condition. Then the powers of the adjoint evidently decompose into tensors with an equal number of upper and lower indexes, each symmetrized in correspondence with some Young diagram, with the additional condition that the contraction of any upper index with any lower index should be zero. This description of irreps can be transformed into a description by Dynkin labels, used in the above, with the help of completely antisymmetric tensor of $N$-th rank. Due to the equality of numbers of upper and lower indexes, i.e. the equality of the areas of these two Young diagrams, at sufficiently large  (with respect to $k$) $N$  these are exactly representations satisfying (\ref{Main2}). 

Consequently, our result shows that all representations appearing in the decomposition of powers of the adjoint representation of $su(N)$, have Casimir eigenvalues that are linear over $N$. This means that these eigenvalues can be, following Vogel's Universal Lie Algebra approach \cite{V0},  represented in the universal form:

\begin{eqnarray}
	C=\alpha x+\beta y + \gamma z
\end{eqnarray}
since universal Vogel's parameters of $su(N)$ for the minimal normalization of roots are given by $\alpha =-2, \beta=2, \gamma=N$ \cite{V0,LM}. See \cite{V0,MSV,AM19} for universal formulae for Casimir, and higher Casimir operators eigenvalues.

This supports the hypothesis of \cite{AIKM} that the $k$-th power of the adjoint can be universally decomposed into Casimir eigenspaces for any $k$.

\section{Acknowledgments}
I'm grateful to the organizers and participants, especially to A. Isaev and S. Krivonos, of the workshop "Universal description of Lie algebras, Vogel's theory and applications", Dubna, April 2025,  for inspiring talks and discussions. 

The work is partially supported by the Science Committee of the Ministry of Science  and Education of the Republic of Armenia under contracts   21AG-1C060 and 24WS-1C031.


\begin{thebibliography}{99}

\bibitem{BR}
A.O. Barut, R. Raczka, Theory of Group Representations and Applications, World Scientific Publ., 1986.

\bibitem{Difr}   di Francesco, P., Mathieu, P. and Senechal, D. (1997) Conformal Field Theory, Springer New York, 1997, 
https://doi.org/10.1007/978-1-4612-2256-9

\bibitem{GG}   
Morikuni Goto and Frank D. Grosshans, Semisimple Lie algebras, Lecture Notes in Pure and Applied Mathematics, V 38, 1978, https://doi.org/10.1201/9781003071778 

\bibitem{Del}
P. Deligne,  La s\'erie exceptionnelle des groupes de Lie, C. R. Acad. Sci. {\bf 322} (1996), 321-326.

\bibitem{V0}
P. Vogel,  The Universal Lie algebra, Preprint (1999), https://webusers.imj-prg.fr/\~{}pierre.vogel/grenoble-99b.pdf

\bibitem{LM} J.M. Landsberg, L. Manivel, A universal dimension formula for complex simple Lie algebra,  Adv. Math. {\bf 201} (2006), 379-407, https://doi.org/10.1016/j.aim.2005.02.007

\bibitem{Cohen} A.M.~Cohen, R.~de~Man, Computational evidence for Deligne's conjecture regarding exceptional Lie groups, C.R.~Acad.~Sci.~Paris 322 (1996) p.427.

\bibitem{AIKM} Maneh Avetisyan, Alexey Isaev, Sergey Krivonos and Ruben Mkrtchyan, The uniform structure of $\mathfrak{g}^{\otimes 4}$, Russian Journal of Mathematical Physics, 2024, Vol 31, No 3, pp 379-338, https://doi.org/10.1134/S1061920824030038

\bibitem{MSV}
R.L. Mkrtchyan, A.N. Sergeev, A.P. Veselov,  Casimir values for universal Lie algebra, arXiv:1105.0115,  Journ. Math.Phys. 53, 102106 (2012).   https://doi.org/10.1063/1.4757763

\bibitem{AM19}	
M.Y. Avetisyan, On Universal Eigenvalues of Casimir Operator,
arXiv:1908.08794, 
Phys. Part. Nucl. Lett. 17(5), pp 779-783 (2020), 
doi:10.1134/S1547477120050039




\end{thebibliography}
\end{document}